\documentclass[aps,pre,floats,twocolumn,showpacs,superscriptaddress,reprint]{revtex4-1}
\usepackage{graphicx,epsfig}
\usepackage{graphics,dcolumn,bm,epic, eepic,fleqn,float}
\usepackage{amssymb,amsmath,amsfonts,multirow,rotate,color,xcolor}
\usepackage{soul}
\usepackage{url}

\usepackage{subfigure}

\begin{document}
\title{A place-focused model for social networks in cities}
\author{Chlo\"e Brown}
\affiliation{Computer Laboratory, University of Cambridge, Cambridge (UK)}

\author{Anastasios Noulas}
\affiliation{Computer Laboratory, University of Cambridge, Cambridge (UK)}

\author{Cecilia Mascolo}
\affiliation{Computer Laboratory, University of Cambridge, Cambridge (UK)}

\author{Vincent Blondel}
\affiliation{Universit\'{e} Catholique de Louvain, Louvain-La-Neuve (Belgium)}
\date{\today}

\begin{abstract}The focused organization theory of social ties proposes that the structure of human social networks can be arranged around extra-network foci, which can include shared physical spaces such as homes, workplaces, restaurants, and so on. Until now, this has been difficult to investigate on a large scale, but the huge volume of data available from online location-based social services now makes it possible to examine the friendships and mobility of many thousands of people, and to investigate the relationship between meetings at places and the structure of the social network. 
\newline In this paper, we analyze a large dataset from Foursquare, the most popular online location-based social network.  We examine the properties of city-based social networks, finding that they have common structural properties, and that the category of place where two people meet has very strong influence on the likelihood of their being friends. Inspired by these observations in combination with the focused organization theory, we then present a model to generate city-level social networks, and show that it produces networks with the structural properties seen in empirical data.
  \end{abstract}

\maketitle
\section{Introduction}
\label{sec:intro}
It is an intuitive idea that social relationships between people arise out of, and are reflected by, meetings and shared activities in common spaces. Scott Feld's theory of the \textit{focused organization} of social ties posits that friendships form between individuals whose interactions are organized around extra-network foci, which can include physical places. In his 1981 paper~\cite{Feld81:Focused}, Feld presents this theory and discusses how commonly observed structural properties of social networks could result.

Empirical investigation of such theories has traditionally been difficult and time-consuming, requiring interviews with, and observation of, necessarily small groups of people. Large-scale analysis has therefore been impossible. However, the recent widespread adoption of location-based online social services has provided us with a huge volume of data both about the structure of people's social networks, as described by social ties explicitly declared by users of these services, and about their activities and meetings at places in their local environment, thanks to the location-sharing dimension. We now therefore have an unprecedented opportunity to investigate the role that places may play in the structure of social networks, on a scale not previously feasible. In addition, the semantic information about places available in location-based online social services allows us to investigate the relationship between the types of places where people meet and the likelihood that those people are friends.

In this work, we study a large dataset from Foursquare, which is the most popular location-based online social network, used by over 35 million people worldwide~\cite{Foursquare12:Reinventing}. We analyze the social and spatial properties of social networks in cities, and present a model for a place-based social network based on our observations from the empirical data, in combination with Feld's focused organization theory. We then show that the model produces networks with the structural properties expected of social networks, as well as preserving the popularity distribution of places in the city and the spans of the sets of places that people visit.

In more detail, our work makes the following contributions:
\begin{itemize}
\item We first define and analyze place-based social networks at the city scale, to answer the question: \textit{what do intra-city social networks look like, and do they have common structural characteristics}? We show that the city-based social networks in the Foursquare dataset have the structural properties observed by computational social scientists studying real-world social networks, namely: a power-law degree distribution, small-world properties (high clustering and small diameter), and strong community structure. While the global properties of online and offline social networks have been analyzed previously, our work is the first to examine and compare the structures of place-based social networks \textit{within different cities} and to show these common structural properties.
\item We then address the question: \textit{is this large location-based social dataset consistent with the theory of focused tie organization}? Exploiting the combination of social information and specific semantic information available in the Foursquare dataset, we are able to examine the relationship between the category of a place where people meet and the probability of friendship. The type of a place where people meet has a strong influence on the likelihood that they are friends, and that there is evidence for the existence of intra-place triads, resulting in clustering of social ties around places in the city.
\item Inspired by these observations in combination with the focused organization theory, we present a model for a city-level social network, based around meetings at places, and show that when simulated this model is able to produce networks with the structural properties observed in real social networks, while preserving the popularity distribution of places in the city and the span distribution of sets of places visited by individuals. The fact that this model is able to reproduce empirically observed social network features is consistent with the idea that places can act as foci for friendships.
\end{itemize}

Our work has intrinsic interest, as we investigate an area largely unexplored, namely, that of the structural similarities between social networks at the city scale within different cities, and demonstrate that the networks in different cities show striking similarities. Furthermore, our model demonstrates that the focused organization of social ties, with places as foci, results in networks with the structural features commonly observed in social networks. 

From a practical perspective, the observation that that the type of a place where people meet strongly affects the probability of friendship could be useful to online location-based social services such as Foursquare. For example, one important application in location-based social networks is the recommendation to users of venues they might want to visit~\cite{Berjani11:Recommendation,Long12:Exploring,Noulas12:Random}. The fact that that the type of place where people meet has a strong influence on friendship suggests that different recommendations would be appropriate depending on the other users present.

There are also potential applications in the development of smarter privacy controls in location-based online social networks: Page et al.~found that people's concerns about privacy in location-sharing services center around the desire to preserve one's existing offline relationship boundaries \cite{Page12:Disturb}, and our observations suggest that these boundaries might be reflected in the types of places where friends meet (for example, closer friends at homes and nightlife spots, less close acquaintances only at professional venues or transport spots). Use of this information could enable services such as Foursquare to adjust the default audience of a check-in, for example, based on relationship semantics inferred using meeting places.

\section{Related work}
\label{sec:related}
The structure of social networks has been well-studied by sociologists and by computational social scientists, and such networks are known commonly to exhibit some particular structural properties. For example, social networks usually have \textit{small-world} properties, that is, small diameter and high clustering~\cite{Pool79:Contacts,Watts99:Networks}, and show a power-law degree distribution~\cite{Barabasi99:Emergence}. Stanley Milgram showed, in his famous study, that social networks tend to have short average path lengths between individuals, finding an average distance of about six hops in the network he studied~\cite{Travers69:Experimental}. Social networks also tend to show strong community structure: the nodes of the network are arranged into tightly-knit groups with their members densely connected, and fewer or looser connections between these groups of friends~\cite{Girvan02:Community}.

In the past decade, the growth in popularity of \textit{online} social networks has enabled the examination of social network structure at a previously impossible scale, and these networks have been shown generally to have the same structural properties as their offline counterparts~\cite{Kumar10:Structure,Mislove07:Measurement,Zhao12:Multiscale}. Location-based social networks in particular have been the subject of much recent research, and many studies have shown that being close geographically and visiting the same places is a strong indicator for being friends in the online location-based social network~\cite{Chang11:Location,Cho11:Friendship,Crandall10:Inferring,Scellato11:Sociospatial}.

The significance of different kinds of places and their associated probability that two people meeting there are friends was discussed by Cranshaw et al.~\cite{Cranshaw10:Bridging}, who developed a metric based on entropy to characterize places according to the diversity of the population of visitors. However, they had a much smaller sample of 489 users than we use here, and they identified locations as cells in a grid, rather than by using specific venues with their associated information available in Foursquare. Other work that has discussed similar concepts includes the studies of urban Bluetooth encounters by Kostakos et al.~\cite{Kostakos09:Understanding,Kostakos10:Brief}, which observed the differences between people's mobility in different places around the city, for example, in pubs and in the street. However, this work only examined the city of Bath, UK, while we have shown similarities across different cities.

Modeling network structure has also long been a highly active research topic, with a huge variety of models being presented over the years to reproduce various empirically observed properties through mechanisms such as preferential attachment, triadic closure, and community association~\cite{Barabasi99:Emergence,Romero10:Directed,Leskovec05:Graphs}. The work closest to that we present here is the body of research concerning affiliation networks, that is, social networks where agents are associated with societies. In this case, the type of place-based city network that we examine can be seen as one where users are associated with places. Many models have been presented that aim to achieve explicit group formation alongside the social network structure~\cite{Lattanzi09:Affiliation,Yang12:Community,Zheleva09:Co-evolution}. Our work differs in that we are not here trying to model well-defined groups in parallel with the social network, but rather we simply use the association of users with places in network generation, in a way inspired by the focused organization theory of social ties.

Most recently, Allamanis et al.~\cite{Allamanis12:Evolution} presented a model for the evolution of specifically location-based social networks. While this model is similar to that we present here in that it takes into account place information in the placement of social ties, they seek to model something fundamentally different. Our work is concerned with creating a place-based network where social ties are reflected in actual meetings between people, not the online social network that commonly contains some ties between people who have never met.

\section{Empirical analysis}
\label{sec:analysis}
In this section, we will describe the Foursquare dataset and explain how we define the city-level social networks. We will then address the question: \textit{what do intra-city social networks look like, and do they have common structural characteristics}? We show that the city-level social networks for different cities not only have similar structure, but also that they have the known properties of general social networks, namely: a power-law degree distribution, small-world properties (high clustering and short paths), and strong community structure.

Furthermore, since the city-level networks have the structure expected of social networks in the real world, we can use the place information from the Foursquare dataset to address the question: \textit{is this large location- based social dataset consistent with the theory of focused tie organization}? The idea is that we can investigate on a large scale whether the idea of focused tie organization, with places acting as the foci, could play a part in the structure of these networks. To this end, we analyze the spatial properties of the social networks, in order to inform our definition of a model for place-focused social networks.

\subsection{Dataset description}
The dataset was collected from Foursquare, which is an online location-based social service where users \textit{check in} to their current location using a mobile application, and share these check-ins with their friends. We collected all of the check-ins from Foursquare posted on Twitter during the 10 months between November 2010 and September 2011; it is estimated that this is 20-25\% of Foursquare check-ins made during this period.

The dataset therefore consists of check-ins having the form (\textit{userID, placeID, timestamp}). We also have the Twitter friend lists of the users concerned, which we use to define a social network, and venue information downloaded directly from Foursquare. 

For every venue, we have the venue ID, venue name, latitude, longitude, and a category giving some indication of the semantics of the place: this is one of the top-level categories provided by Foursquare, namely: Arts and Entertainment, College and University, Food and Drink, Nightlife Spot, Outdoors and Recreation, Professional and Other Places, Residence, Shop and Service, and Travel and Transport. 

In the following, we present the data from five large US cities for which we have a large number of users and check-ins: Atlanta, Boston, Chicago, Minneapolis and Seattle. The numbers of users, venues and check-ins in the dataset for each city are shown in Table \ref{tab:datasets}.
\begin{table*}
\centering
\begin{tabular}{|c||c|c|c|}
\hline
City & Users & Venues & Check-ins  \\
\hline
\hline
Atlanta  & 28,275 & 18,270 & 368,608 \\
\hline
Boston & 23,579 & 13,243 & 296,150 \\
\hline
Chicago  & 42,791 & 33,261 & 715,652 \\
\hline
Minneapolis  & 13,396 & 12,696 & 235,793 \\
\hline
Seattle  & 16,205 & 15,051 & 260,023 \\
\hline
\end{tabular}
\caption{Number of users, number of venues, and number of check-ins in the dataset, for each of the five cities.} 
    \label{tab:datasets}
\end{table*}

\subsection{City social networks}
We define a place-based social network for each city, using the Twitter friend lists of the users in the dataset. Since we want to study specifically the social networks within cities grounded in physical space, we constructed a social network for each city by considering users to be friends when each is in the friend list of the other, and the connected users have checked in to at least one of the same places on Foursquare. We require the tie to be reciprocated in Twitter's directed social graph to remove ties where, for example, many users may follow a celebrity without actually knowing them, but the celebrity does not follow all of their followers in return. A reciprocal tie must be approved by both users, and so better represents some definition of friendship.  The place requirement is imposed in order to ground the network in physical space.

Formally, for each city, given:
\begin{enumerate}
	\item The set of $n$ users: $U = \{u_1, u_2, u_3, \ldots, u_n\}$
	\item The Twitter friend lists for each user $u_i \in U$: 
	
	$F_i = \{ u_j$  $|$ $u_i$ follows $u_j $ on Twitter$\}$
	\item The sets of venues for each user $u_i \in U$:
	
	$V_i = \{ v$ $|$ user $i$ has checked in to venue $v$ in the city$\}$
\end{enumerate}
we construct the social graph $G(U, E)$ where the $n$ nodes $U$ of the graph represent the users and the graph $G$ has an undirected edge $(u_i, u_j) $in the set of edges $E$ whenever:
\begin{enumerate}
	\item $u_i \in F_j$ and $u_j \in F_i$, i.e. the users both follow one another on Twitter, and
	\item $| V_i \cap V_j | > 0$, i.e. the users have both checked in to at least one of the same places.
\end{enumerate}

\subsection{Structural properties}
\begin{table*}
\centering
\begin{tabular}{|c||c|c|c|c|c|c|c|c|c|}
\hline
City & $N$ & $K$ & $N_{GC}$ & $C$ & $C_r$ &$d$ & $d_r$ &$Q$ & $Q_r$  \\
\hline
\hline
Atlanta  & 13,011 & 46,756  & 11,476 & 0.16 & 0.0006 & 4.6 & 4.6 & 0.53 & 0.17\\
\hline
Boston & 10,478 & 41,505  & 8,816 & 0.17 & 0.0010 & 4.3 & 4.0 & 0.45 & 0.15\\
\hline
Chicago  & 19,931& 84,778  & 17,287 & 0.16 & 0.0004 & 4.6 & 4.9  & 0.47 & 0.14\\
\hline
Minneapolis  & 6,499 & 30,640 & 5,914 & 0.18 & 0.0016 & 4.4  & 4.2 & 0.41 & 0.12\\
\hline
Seattle  & 7,445 & 28,466 & 6,392 & 0.18 & 0.0008 & 4.4  & 4.6 & 0.46 & 0.16\\
\hline
\end{tabular}
\caption{Numbers of nodes $N$ and edges $K$, nodes $N_{GC}$ in the giant connected component, average clustering coefficient $C$, average shortest path length $d$, and modularity $Q$, average clustering coefficient $C_r$, shortest path length $d_r$, and modularity $Q_r$ in a random network with the same number of nodes and edges, for each of the five cities. The high clustering coefficient and similar shortest path length with respect to the random network shows that these social networks have small-world properties.} 
    \label{tab:clustering}
\end{table*}
We first analyze the structures of the city-level social networks and confirm that they exhibit well-known characteristics of social networks: a power-law degree distribution, small-world properties (high clustering with respect to a random network, and small shortest path lengths) and strong community structure.
\begin{figure*}
        \centering
        \includegraphics[width=0.7\textwidth]{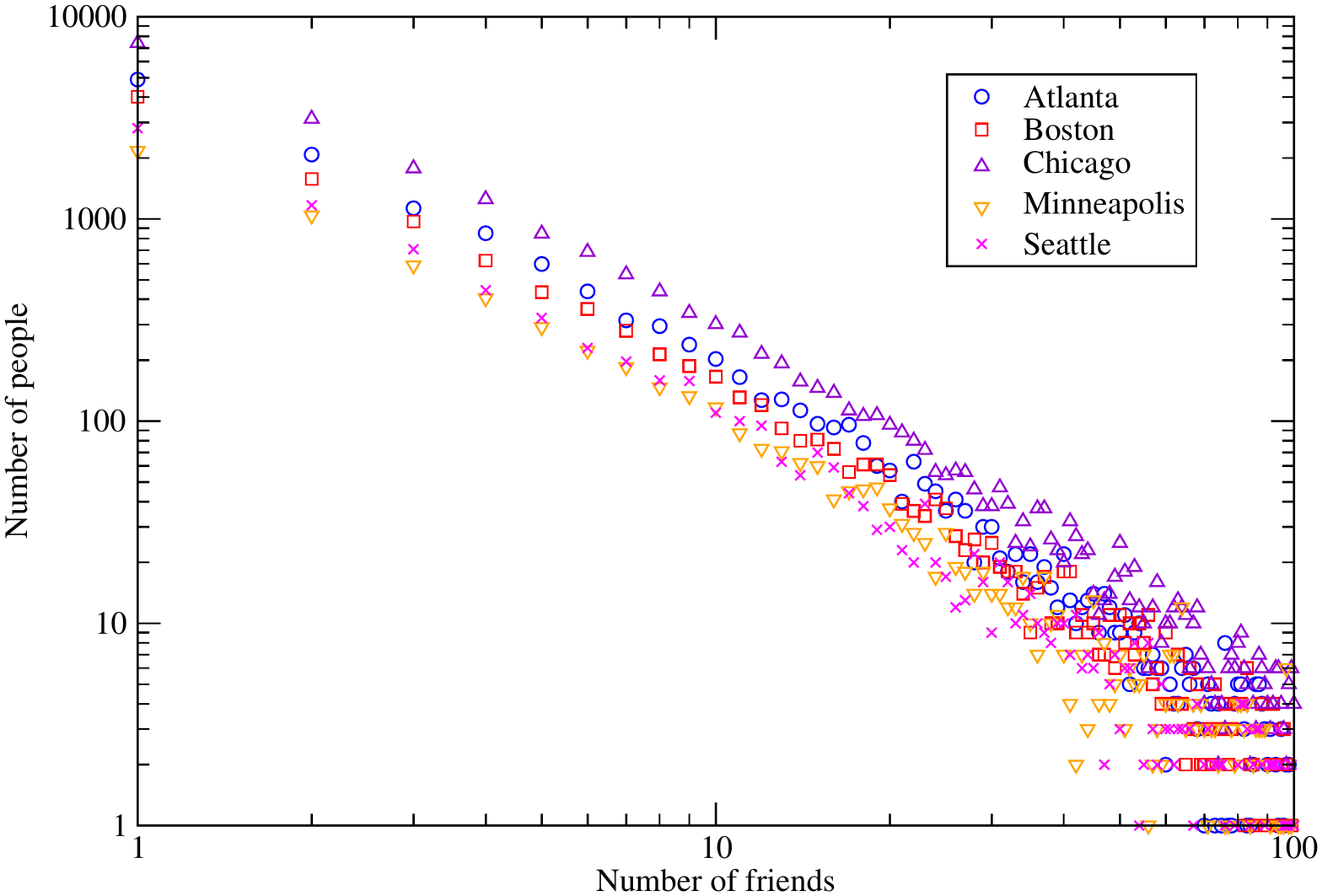}
        \caption{Degree distributions in the city-level social graphs; each distribution plotted on a log-log scale presents a linear functional form similar to a power-law, with many people having few friends and a few people having very many friends. While the exact numbers of people having each degree differ according to the number of people in the dataset for each city, the exponents of the distributions, reflected in the slopes of the graphs, are remarkably similar across the cities. The average exponent across the 50 cities in the Foursquare dataset with the most check-ins is 2.76.}\label{fig:degrees}
\end{figure*}
\subsubsection{Degree distribution}
Many networks, from those in biological systems such as metabolic networks, to technological structures such as the topology of the Internet and the page structure of the World Wide Web, exhibit a power-law degree distribution, and social networks are known also to have this property~\cite{Barabasi99:Emergence,Girvan02:Community,Amaral00:Classes,Wagner01:Small}. Essentially, this means that the number of others to whom an individual in the network is connected is distributed as a power-law, so that while most have a small number of ties, there are a few hub nodes with a large number of connections. 

Figure \ref{fig:degrees} shows the degree distributions of the social networks for the cities we observe in our dataset. We can see that indeed, the degree distributions of the city-level networks in our dataset resemble power-laws.
While the exact numbers of people having each degree differ according to the number of people in the dataset for each city, the exponents of the distributions, reflected in the slopes of the graphs, are remarkably similar across the cities. Using the methodology presented in \cite{Clauset09:Power}, we have confirmed the power-law distribution and measured an average exponent of 2.76 across the 50 cities in the Foursquare dataset with the most check-ins.

\subsubsection{Clustering}
Another commonly observed property of social networks is a relatively high level of clustering, compared to the level seen in a random graph~\cite{Watts99:Networks}. In terms of social relationships, this corresponds to the fact that many of an individual's friends are likely to be friends with one another. The level of clustering in a graph can be measured by the clustering coefficient. The clustering coefficient $C$ of a node with $N$ neighbors is defined as the number of links between these $N$ neighbors, divided by the number of possible links that could exist between the neighbors, i.e. 
\begin{equation}
\frac{2N}{N(N-1)}
\end{equation}
The clustering coefficient $C$ of a graph is then defined to be the mean clustering coefficient of all its nodes. Table \ref{tab:clustering} shows the clustering coefficients observed in the networks for our five cities under analysis. We can see that the values are consistent across all the five networks, being between 0.1 and 0.2. This is much higher, on the order of thousands of times, than the level observed in random graphs with the same numbers of nodes and edges, as shown in the table.

\subsubsection{Average shortest path length}
Social networks are known commonly to have a low average shortest path length. A shortest path from one node $m$ to another $n$ is defined to be the smallest number of steps that are needed to reach $n$, starting at $m$ and travelling along edges in the graph. The average shortest path length is then defined to be the mean value over all pairs in the graph. This quantity is defined only for connected graphs, and so for our networks we consider only the giant component that always contains at least 80\% of the nodes in the graph, the presence of which is indeed another characteristic commonly seen in social networks~\cite{Kumar10:Structure}. The average shortest path lengths $d$ in the giant components of the social networks in the dataset can be seen in Table \ref{tab:clustering}, and can be seen to be comparable to those in random graphs, which together with the high clustering denotes so-called small-world networks~\cite{Watts99:Networks}. 

\subsubsection{Community structure}
The final prominent structural feature of social networks we consider here is community structure. Social networks tend to exhibit strong community structure, that is, the nodes are arranged into groups tightly connected by many social ties between members of the group, and these groups are more loosely interconnected to make up the entire network~\cite{Girvan02:Community,Fortunato10:Community,Onnela12:Taxonomies}. One measure of the strength of community structure in a network is the modularity $Q$ of a partition of a network into its communities. Values of 0.3 or above are generally considered high, and indicative of the kind of community structure commonly seen in various biological, technological and social networks~\cite{Newman06:Modularity}.

We partition our social networks into communities using the Louvain algorithm~\cite{Blondel08:Fast}, a popular community detection method that scales well for large networks, and measure the modularity of the partition in each case. The values of $Q$ shown in Table \ref{tab:clustering} show that the place-based social networks in the five cities do indeed have strong community structure.

\subsection{Spatial properties}
We now investigate in more detail the spatial properties of the social networks in the dataset, to try to gain insight into the nature of place-based friendships. We study the popularity distribution of places in the dataset and find that it resembles a power-law, similarly to the degree distribution of the social network. We further analyze triangles in the network and find that most triads have at least one common place, suggesting that some intra-place triadic closure mechanism could be at work. Finally we study the likelihood of friendship between colocated people given the category of that place, and show that the type of place where people meet has a strong influence on the probability that they are friends. 
\begin{figure*}
                \centering
                \includegraphics[width=0.7\textwidth]{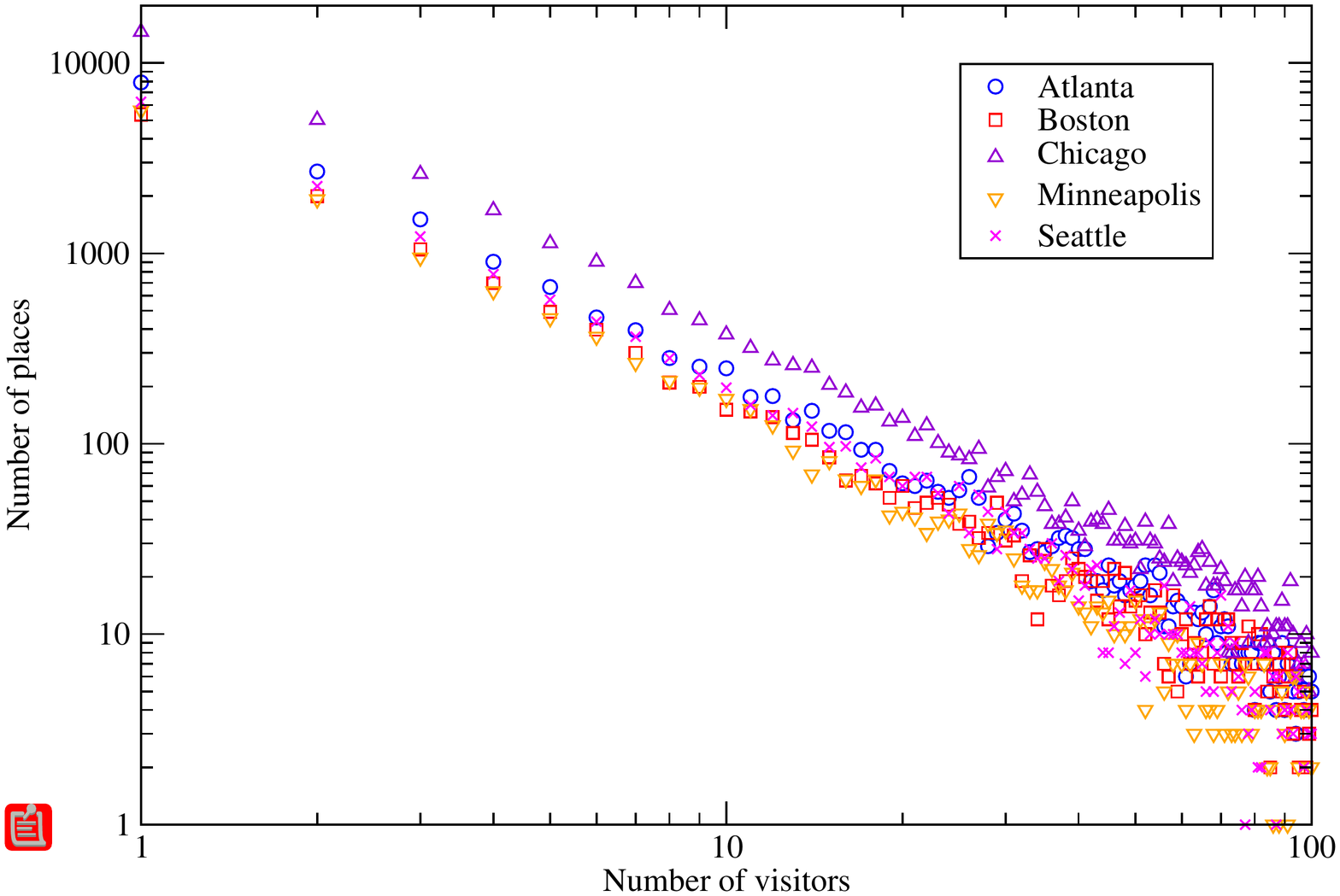}
        \caption{Numbers of people checking in at a place in the dataset. The place popularity distribution on a log-log scale presents a linear functional form similar to a power-law, with many places having few visitors and yet a few places being extremely popular. The mean exponent for the power-law distributions we measured across the 50 cities with the most check-ins was 1.87.}\label{fig:place_popularity}
\end{figure*}

\subsubsection{Place popularity}
Figure \ref{fig:place_popularity} shows the popularity of places in each city, defined to be the number of unique people who have checked in at that place in the dataset. The place popularity distribution on a log-log scale presents a linear functional form similar to a power-law, with many places having few visitors and yet a few places being extremely popular. The mean exponent for the power-law distributions we measured across the 50 cities with the most check-ins was 1.87.

\subsubsection{Proportions of triads having one common place}
We consider the groups of three friends $u_1$, $u_2$ and $u_3$ where the ties $(u_1, u_2)$, $(u_2, u_3)$ and $(u_3, u_1)$ all exist in the social graph, and examine the proportion of such triangles where at least one place has been visited by all three people. Table \ref{tab:triangles} shows the proportions for each of the five cities.
\begin{table*}
\centering
\begin{tabular}{|c||c|}
\hline
City & Proportion of triangles  \\
\hline
\hline
Atlanta  & 0.90\\
\hline
Boston & 0.81\\
\hline
Chicago  & 0.80\\
\hline
Minneapolis  & 0.71\\
\hline
Seattle  & 0.84\\
\hline
\end{tabular}
\caption{Proportions of social triangles where all three friends have at least one common place. In each city, more than 70\% of triangles in the place-based social network are such that the same place is shared between each pair of friends in the triangle.} 
    \label{tab:triangles}
\end{table*}
The vast majority (at  least 70\%) of triangles are such that each pair of friends making up a triad shares at least one place, indicating the presence of clustering around common places. It would therefore seem promising to consider including triangles made up of people visiting the same places in our model.

\subsubsection{Types of places and likelihood of friendship}
We now investigate the potential for meetings at various places around the city to foster or to reinforce friendships. To this end, we make use of the categories provided by Foursquare for each venue in their database. Each place has one category assigned by Foursquare, and the possible categories for a places are: Arts and Entertainment, College and University, Food, Nightlife Spot, Outdoors and Recreation, Professional and Other Places, Residence, Shop and Service, and Travel and Transport.

We consider pairs of users who have been colocated at venues, that is, they have not only visited the same place but they have also had the chance to meet there through being there at the same time. Since check-ins are identified by a single timestamp, we consider users to have been colocated at a venue if they both checked in there within a 1 hour time window. We then compute the probability that a pair have a tie in the social network for a city, given that they have been colocated at a place.
\begin{figure*}
        \centering
        \includegraphics[width=0.7\textwidth]{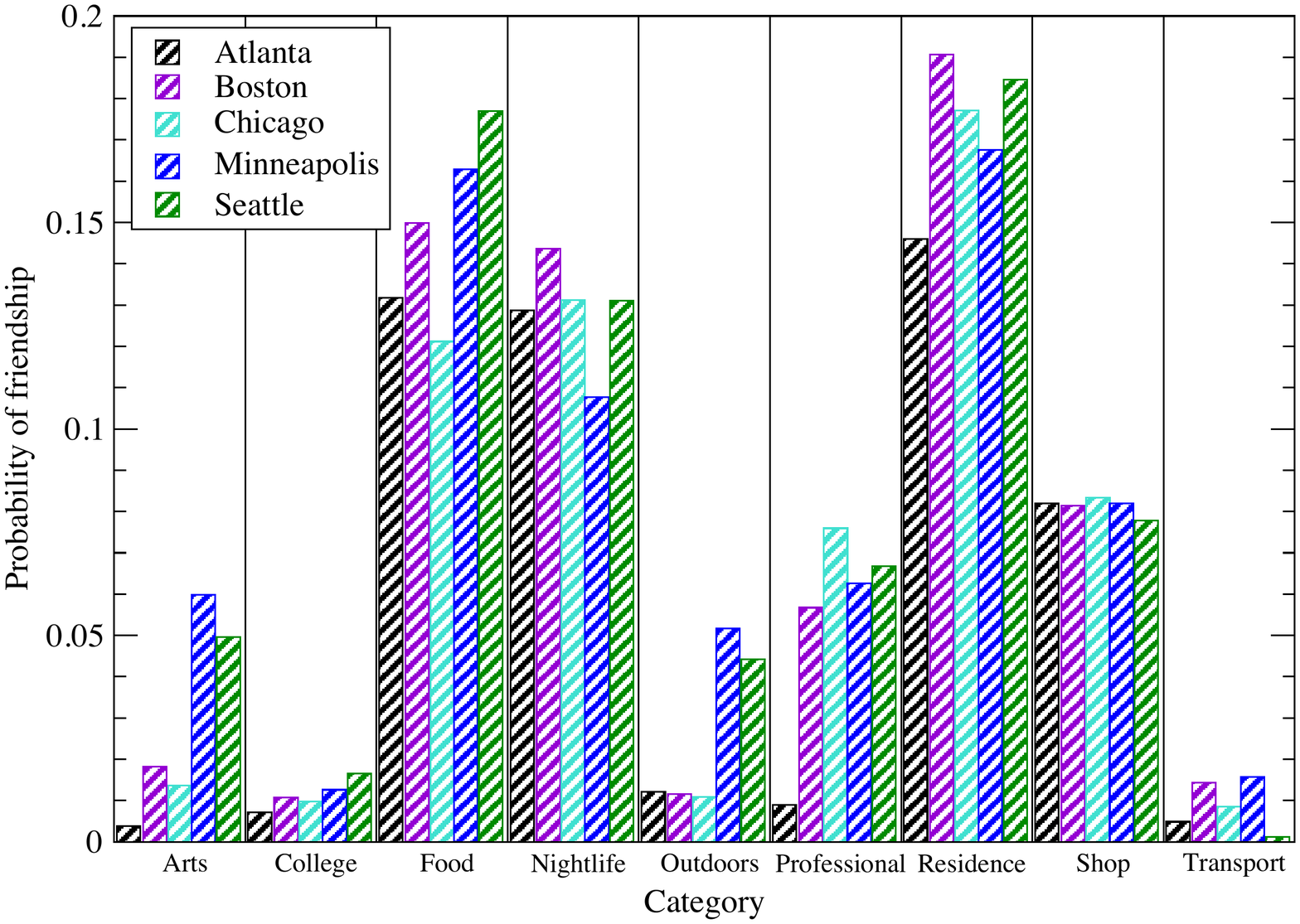}
        \caption{Probability of friendship between pairs colocated at a place, by Foursquare category. Categories of places have one of three types: `social' places, where people tend to meet with their friends, `semi-social' categories, where some meetings between friends take place but where there are also commonly meetings with strangers, and generally `non-social' places, where people tend to meet with strangers over friends to a greater degree than in other kinds of places.}\label{fig:colocs}
\end{figure*}

Figure \ref{fig:colocs} shows these probabilities for each city. We can see that some categories (Food, Nightlife and Residence) have a far greater likelihood that colocated pairs are friends than the others do, which suggests that some places have a greater potential for fostering and for reinforcing social ties, or acting as a focus for friendships, than others. Specifically, across the cities, the probability that a pair colocated at a Food, Nightlife Spot, or Residence venue are friends is between 0.1 and 0.2. The probability for the categories Professional and Other Places, and Shop and Service, is lower, being between 0.05 and 0.1, and the other categories generally have a very low probability. 

This divides places into three categories: `social' places, where people tend to meet with their friends, `semi-social' categories, where some meetings between friends take place but where there are also commonly meetings with strangers, and generally `non-social' places, where people tend to meet with strangers over friends to a greater degree than in other kinds of places.

Again, this is consistent with the focused tie organization theory as it suggests that some places have high potential for fostering or for reinforcing friendship, acting as the foci described by the theory. In the next section, we use our observations from the Foursquare dataset to define a model for this type of place-focused network, in order to test computationally whether it results in networks with the known structural characteristics of social networks.

\section{A model for place-focused social networks}
\label{sec:model}
Having analyzed the Foursquare dataset, we now use our observations to define a model for place-focused networks, based around people meeting at places in the city. This represents a large-scale investigation of how the focused organization theory fits with empirical data. Moreover, it provides an opportunity to investigate how the way in which people meet one another at places around the city could affect the structure of the resulting social network, by running the same model without various components, which we will show later.

\subsection{Model description}
Our model uses three main kinds of information about the places in the city:
\begin{enumerate}
	\item The popularity of a place, i.e. the number of users who have checked in there. 
	\item The geographic (latitude and longitude) coordinates of a place.
	\item The semantics of a place and the activities that take place there, as indicated by the Foursquare categories described in the previous section.
\end{enumerate}

These first two pieces of information are relevant to ensure that the mobility of people described by our model is consistent with the mobility patterns that we see in the real world. The information about place categories, on the other hand, is used to define the potential for a place to act as a focus for friendship, given our observations in the previous section. The procedure to generate a social network in the city is defined as follows:
\begin{enumerate}
	\item Begin with the set of $N$ people and $V$ venues in the dataset for the city.
	\item For each person $u$, consider them to have visited $m$ places, where $m$ is sampled from the distribution of places per user in the dataset for the city. Assign to $u$ an initial place from the set $V$, chosen with probability proportional to the popularity of that place.
	\item For each of the $m-1$ additional places that $u$ has visited in the dataset, assign to $u$ another place $v$ from $V$, with probability $qr^{\alpha}$, where $q$ is proportional to the popularity of $v$ in the dataset, and $r$ is inversely proportional to the rank distance of the place from the first place of $u$. Given a set of places $V$ in a city, the probability of a person visiting place $v_1 \in V$ given that they have visited place $v_2 \in V$ is defined to be:
	\begin{equation}
	r \propto \frac{1}{rank_{v_1}(v_2)}
	\end{equation} where 
	\begin{equation}
		rank_{v_1}(v_2) = n+1
	\end{equation} when $n$ is the number of places in the city closer to $v_1$ than $v_2$.
	This method of place assignment is designed to reproduce observed patterns of intra-city human mobility. We choose the exponent $\alpha = 0.84$ in accordance with the study of human mobility in cities~\cite{Noulas12:Tale}, in which the authors found that people's movements within a wide range of cities of varying sizes and densities tend to follow this distribution.
	\item For each place $v$ in $V$, for each pair of people ($u_1, u_2$) who have been assigned place $v$: \begin{enumerate}\item Place a social tie between $u_1$ and $u_2$ with probability
	\begin{equation}
	p = \left\{
	\begin{array}{ll}
		p_{cat}(v)  & \mbox{if } pop(v) \leq 30 \\
		0.001 & \mbox{otherwise}
	\end{array}
	\right.
	\end{equation}
	according to the category of $v$: $p_{cat}(v) = 0.15$ for `social' places (Food, Nightlife Spot, and Residence), $p_{cat}(v) = 0.08$ for `semi-social' places (Professional and Other Places, and Shop and Service), and $p_{cat}(v) = 0.01$ for all other places, in the light of the analysis in the previous section. $pop(v)$ is the popularity of $v$; we discuss the choice of the threshold 30 in the following section.
	\item For each of $u_1$'s existing friends $f$ in the social network who have visited $v$, place a link between $f$ and $u_2$ with probability 0.15, in line with the `social' probability. This implements an intra-place triangle closing mechanism as suggested by the observation in the previous section that most triangles in the networks have one common place shared by all three members.
	\end{enumerate}
\end{enumerate}

\section{Properties of synthetic networks}
\label{sec:evaluation}
We now present the properties of the synthetic networks generated using the procedure described above. We show that the model produces networks with the structural properties observed in the real data, and demonstrate how the model is also able to preserve important spatial properties such as the popularity of places and the dispersion of places visited by mobile users in real traces. The results given for each city are the average of 10 runs of the model, apart from degree distribution, place popularity, and geographic span, for which we choose one distribution randomly from those generated by the 10 runs. The distributions across all of the 10 runs were similar.

We then study the effect of each piece of information by removing each one in turn, running the model, and examining the properties of the resulting networks: first, we analyze the effect of intra-place triadic closure, and then examine the effect of the place information. This information includes distance (place location), place category, and place popularity. Our analysis shows that these place-based features of our model are important for the generated networks to show the properties observed in real social networks. 

\subsection{Full model}
The results of running the full model, incorporating information about distance, place semantics, and including the triangles between people meeting friends of friends, are shown in Figure \ref{fig:synthetic_degrees} and Table \ref{tab:emergent}.

\subsubsection{Structural properties}
\begin{figure*}
        \centering
	\includegraphics[width=0.7\textwidth]{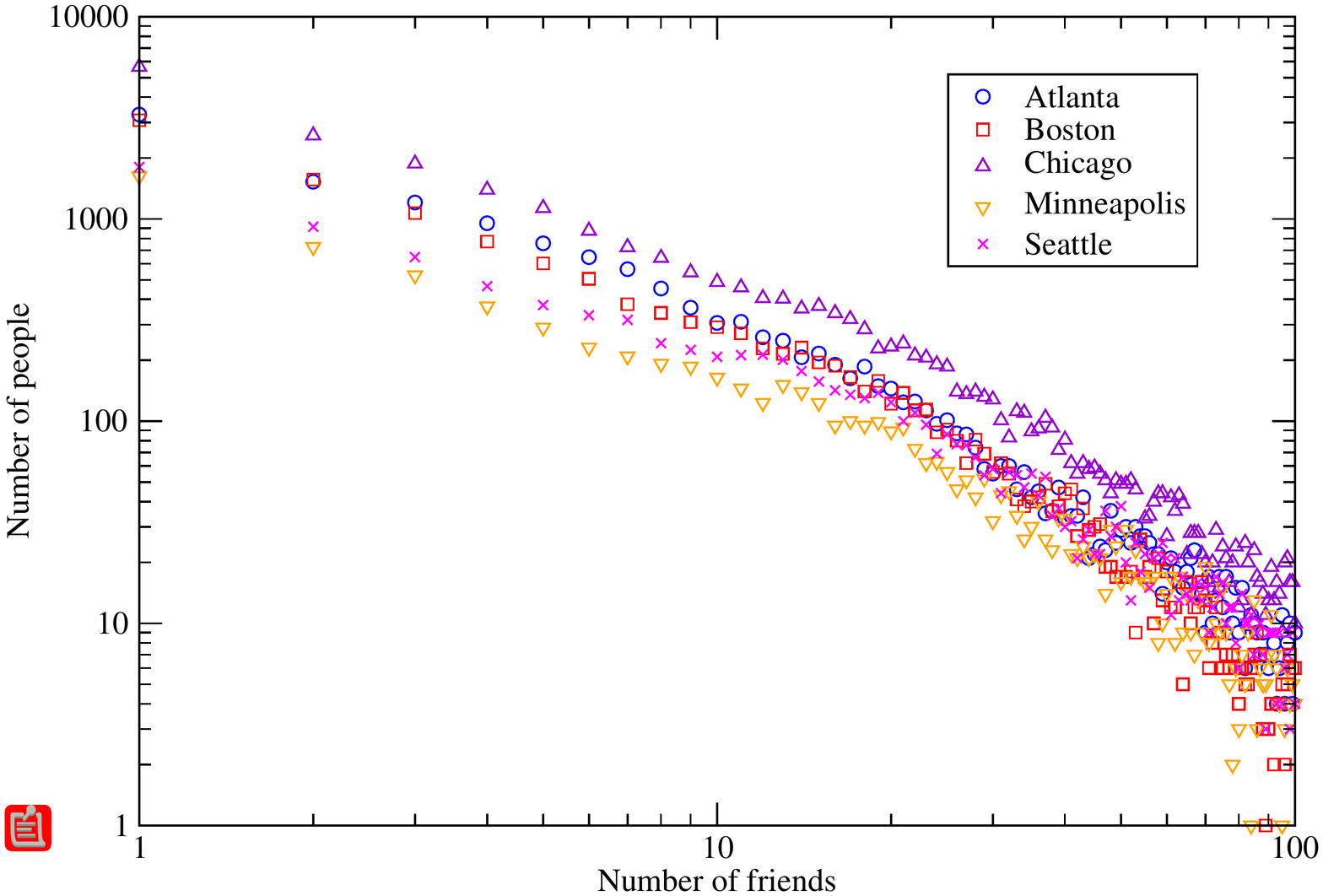}
        \caption{Degree distributions of synthetic graphs. The full model produces networks with a power-law degree distribution, as seen in the empirical data for each of the five cities.}\label{fig:synthetic_degrees}
\end{figure*}

Figure \ref{fig:synthetic_degrees} shows that the full model produces networks with a power-law degree distribution, as seen in the actual social networks. As seen in Table \ref{tab:emergent}, the synthetic graphs also display the structural properties of the empirically observed networks: high clustering (between 0.1 and 0.2), short path lengths (around 4 hops), and strong community structure (as indicated by a modularity value $Q$ higher than 0.3).
\begin{table*}[t]
\begin{centering}
\begin{tabular}{|c|c|c|c|}\hline
Model&$C$&$d$&$Q$\\\hline
Full&0.14&4.0&0.40\\\hline
No distance&0.13&3.8&0.33\\\hline
No categories&0.01&3.8&0.29\\\hline
No triadic closure&0.05&4.0&0.38\\\hline
\end{tabular}
\caption{Properties of the synthetic place-based social networks. The figures shown are averages over the cities. The individual values differ between cities but are close to the average in each case. The full model produces networks with the correct properties, namely high clustering coefficient $C$, low average shortest path length $d$ comparable with that seen in random graphs, and strong community structure as indicated by high modularity $Q > 0.3$.
Removing individual components from the model affects these properties of the generated networks.}
\label{tab:emergent}
\end{centering}
\end{table*}

\subsection{Spatial properties}
Besides the known structural properties of social networks that are present in the generated networks, the important spatial properties of user visits to places within the system are also preserved: the popularity of places, and the distribution of geographic distances between places visited by a user.
\begin{figure*}
	\centering
	\includegraphics[width=2.0\columnwidth]{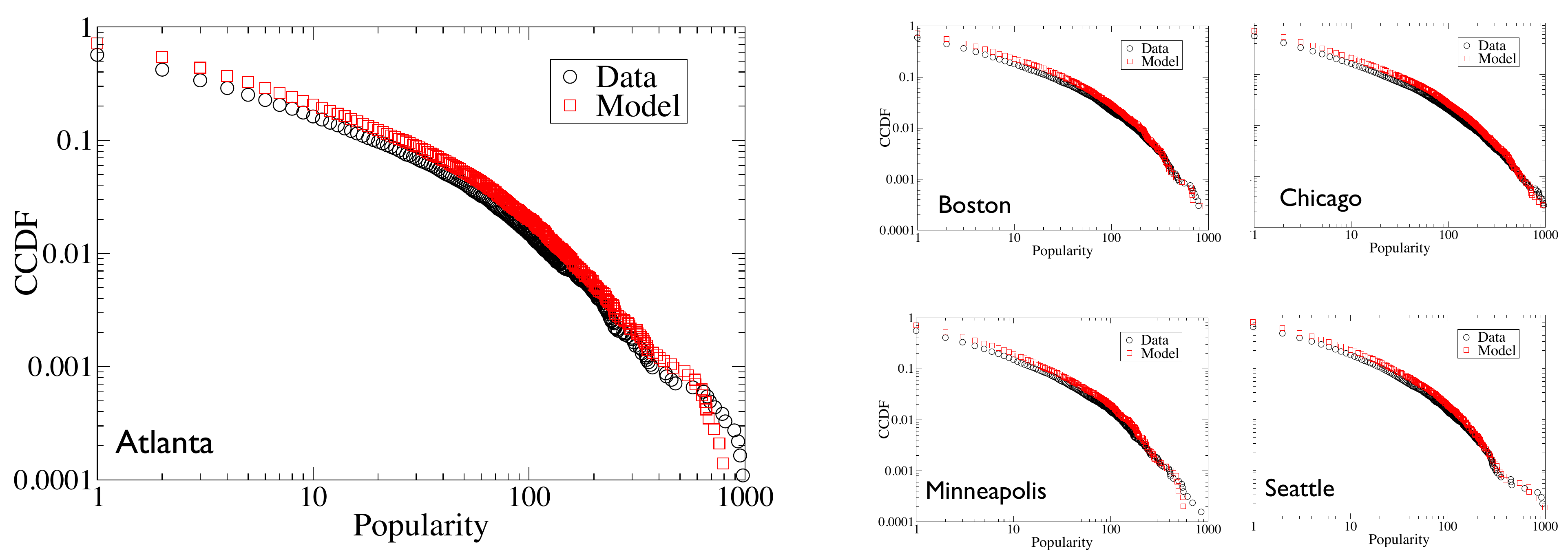}
	\caption{Complementary Cumulative Distribution Function (CCDF) of place popularity, in the model (red squares), and in the real data (black circles), for each of the five cities. The method of assigning places to users according to a combination of place popularity and rank distance is effective to preserve the distribution of popularity of places.}\label{fig:synthetic_popularity}
\end{figure*}
\begin{figure*}
	\centering
	\includegraphics[width=2.0\columnwidth]{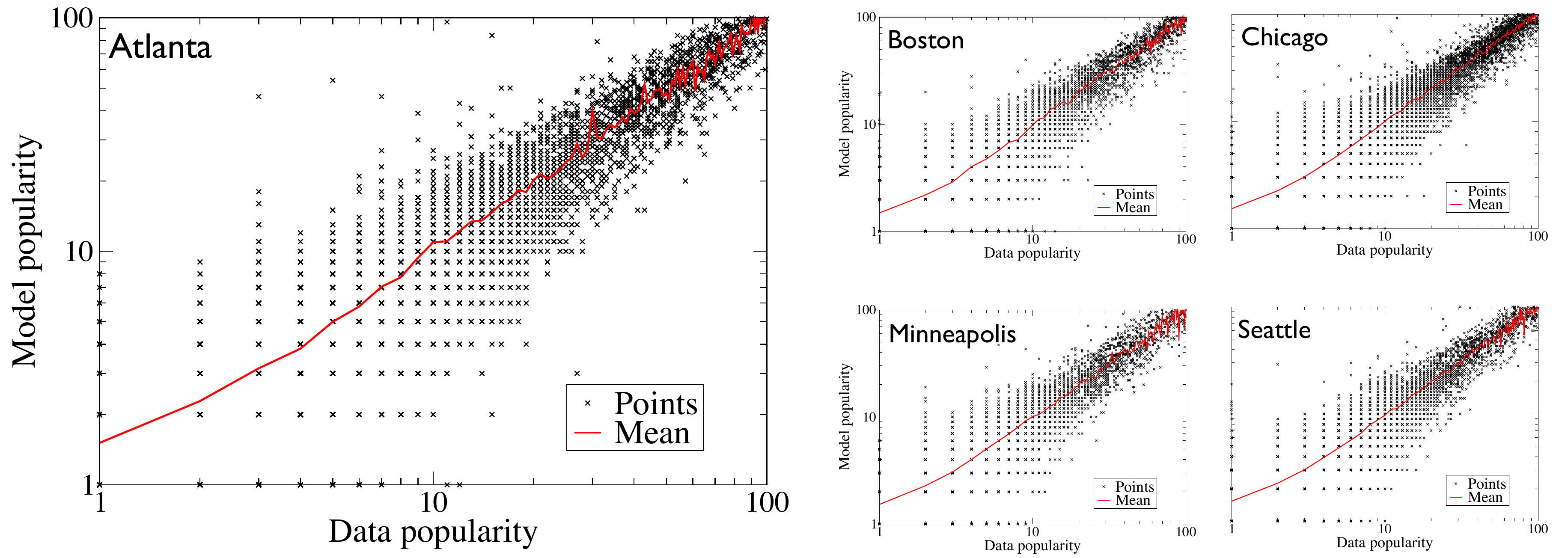}
	\caption{Popularity of individual places in the model and in the real data, for each of the five cities. The popularity of a place is preserved by the model, so places that are unpopular in the empirical data are not disproportionately popular in the assignment of places to users generated by the model.}\label{fig:synthetic_popularity_comparison}
\end{figure*}

\subsubsection{Place popularity}
We consider two measurements of venue popularity in cities. First, in Figure~\ref{fig:synthetic_popularity} we present the CCDF of place popularity distribution as seen in the real data and the assignment of places to users performed by the model, using rank distance and popularity as described in the previous section. For all cities the model approximates the empirical distribution well. The skewed distribution on a log-log scale implies that the original hierarchy of places, where a few places are very popular and many are visited by only a few people, is preserved.

From a social network modeling perspective, we desire this property because the category of a place, used to determine the probability of an intra-place tie, can be related to its popularity (e.g.~an airport or railway station, both in the Transport category, may be very popular but also unlikely to be a focus for friendships). Therefore, we do not wish places that are not very popular in the real data to be extremely popular in the model's assignment of people to places. 

Figure~\ref{fig:synthetic_popularity_comparison} shows a different perspective of place popularity: for every place in the city visited by a given number of users, we plot the corresponding number of users assigned to that place by the model. The red line shows the average modeled popularity of a place with a given empirical popularity; the value for the model assignment closely matches that in the empirical data.

\subsubsection{Geographic span of individual users}
The model also preserves the distribution of geographic span of places visited by individual users. This is a measure of the spread over space of the set of places $P$, with each $p \in P$ having coordinates $(p_x, p_y)$, that a user visits, and is computed as:
\begin{equation*}
span(P) = \frac{\sum_{p \in P}{dist(p, c)}}{|P|}
\end{equation*}
where $c$ is defined to have coordinates $(c_x, c_y)$ where
\begin{equation*}
	c_x = \frac{\sum_{p \in P}{p_x}}{|P|}
\end{equation*}
and
\begin{equation*}
	c_y = \frac{\sum_{p \in P}{p_y}}{|P|}
\end{equation*}
and the function $dist(p_1, p_2)$ returns the great-circle distance between places $p_1$ and $p_2$.
\begin{figure*}
\centering
\includegraphics[width=2.0\columnwidth]{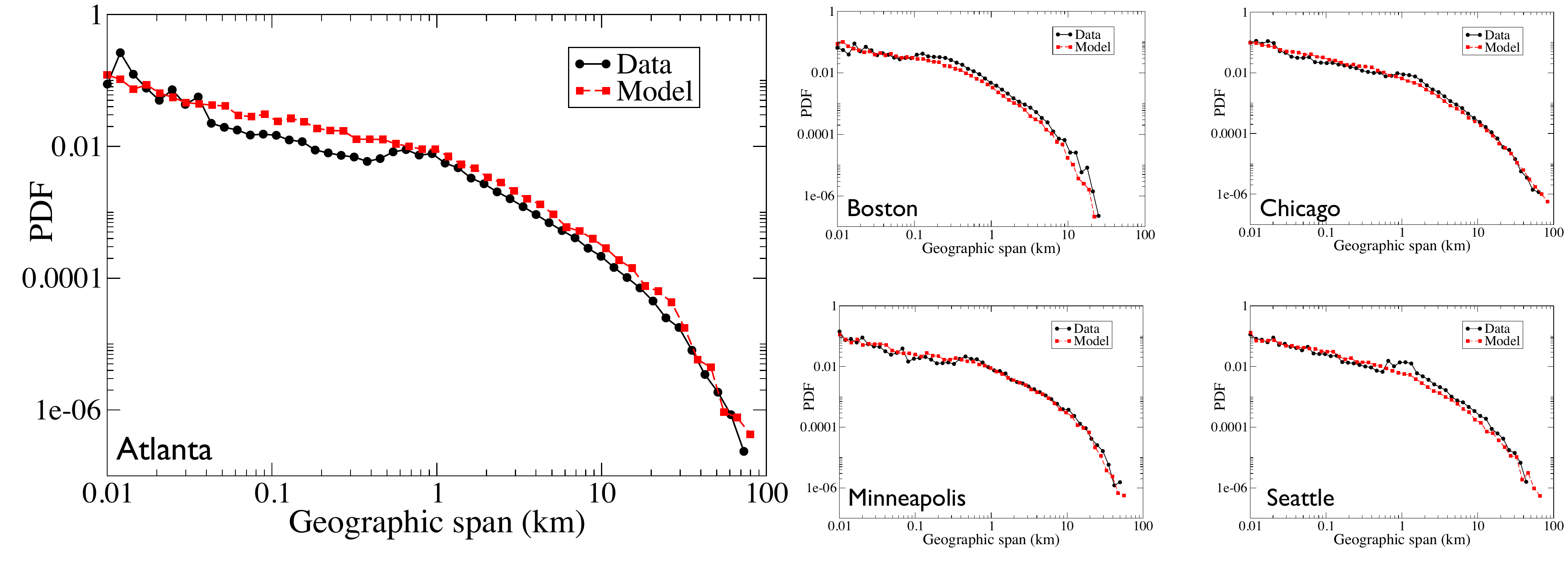}
\caption{Probability Density Function of the geographic span of the places visited by individual users in the model and in the real data, for each of the five cities. The span distribution is generally preserved by the model, meaning that there is the same chance that a user will visit places spread over a certain distance in the city. People are generally more likely to visit places within a small geographic range, with occasional longer trips.}\label{fig:synthetic_span_comparison}
\end{figure*}

The distributions of users' geographic span in each of the five cities are shown in Figure~\ref{fig:synthetic_span_comparison}. The model gives the same distribution of spans as in the empirical data, with users being more likely to visit other venues nearby than places further away. This is in accordance with empirical observations made in studies of human mobility in cities~\cite{Noulas12:Tale}, confirming that the mechanism we have used for assigning places to users does not produce an unrealistic pattern of user visits to places. 

\subsection{Effect of individual components of the model}
We now investigate the effect of each kind of information used by our model on the structure of the generated social networks, by running the model using all of the information except that in question, and examining the properties of the networks that result.

\subsubsection{Model without triadic closure}
We first examine the effect of the triadic closure mechanism on the networks, where when a person has a social tie to another at a common place, they also have a social tie to that other person's friends within that place. Table \ref{tab:emergent} shows that in the absence of the triadic closure mechanism, when social ties are placed only according to the probabilities of connection at given places according to their category, clustering in the graph is unsurprisingly much lower than in the actual networks. Triadic closure is an idea often explained in terms of social balance theory~\cite{Cartwright56:Structural}, in that if an individual has two friends who themselves are not friends, more psychological stress will result than if those two friends are also friends and the relevant triangle exists in the social graph. One can imagine that this effect would be particularly strong if the individuals concerned were actually meeting face to face, such as would be likely when those three individuals are connected by the sharing of a common place.

\subsubsection{Model without distance}
To study the effect of the distance between places, and people's tendency to visit places closer to others that they visit, on the networks, we run the model assigning places to users using only the popularity of the place (instead of using the popularity combined with the rank distance as outlined in the previous section). The table shows that this reduces the modularity value $Q$ of the resulting networks, which implies weaker community structure. This can be understood in terms of people being more likely to visit, and therefore to meet with friends at, places nearby to locations they already visit. When distance is not taken into account in the model, this additional clustering effect based on geography disappears and leaves only clustering at single places produced by triadic closure, which weakens the community structure of the network.

\subsubsection{Model without categories}
We study the effect of the different probabilities of ties between pairs at places of different categories by running the model without the category-based probabilities and instead using in each case a uniform probability that will produce the same number of ties in the resulting network. The table shows that this dramatically reduces the clustering coefficient $C$ of the resulting network, even when the triadic closure mechanism is present. This is because the clusters of social ties tend to be around the more sociable places where friends tend to meet, that have higher probabilities for intra-place ties in the full model. Even when the same number of ties are placed, if these are not focused around `social' places but spread over all the venues with equal probability, the network does not display the high clustering characteristic of a social network. We see also that the community structure of the network is weaker than when the full model is used.

\section{Discussion}
\label{sec:discussion}
The mechanism of assigning people to places based on popularity and forming intra-place ties has much in common with the preferential attachment model~\cite{Barabasi99:Emergence}, which was proposed to produce scale-free networks by means of having new nodes connect to existing nodes with probability proportional to their degree. That is, the graphs grow in a `rich-get-richer' fashion; the more neighbors a node already has, the more likely it is that it acquires more neighbors, which produces graphs with a power-law degree distribution. Our model can be seen as employing a version of preferential attachment to places: more popular places are more likely to attract more people, and therefore more potential pairs of friends. 

However, we have not used only place popularity, but have taken into account the categories of places to determine the likelihood of intra-place ties. We have specified that the probabilities of social tie formation based on the categories of places are used when the places in question have 30 or fewer visitors, and a lower probability for all more popular places. In our experiments, we found that this was necessary in order to be able to obtain networks with the correct community structure, resembling what we see in the real network, where most members of a community share one common place, due to the property that most triangles have one place shared between all three friends. This might intuitively be considered to be a manifestation of the idea of the \textit{constraint} of a place outlined in the 1981 paper proposing the focused organization theory for social tie formation~\cite{Feld81:Focused}. A place with high constraint forces individuals there to interact much and often (for example, a family home) resulting in more or stronger social ties than a place with low constraint where many individuals may go, but will probably not encounter one another, or where visitors to the place do not typically interact with many others (for example, a park). Highly-constraining foci that, as Feld writes, ``create close-knit clusters of various sizes", correspond to the places with a fairly small number of visitors in the city.

Interestingly, this threshold is the same as that found in a study of the geographic span of communities in a mobile phone communication network~\cite{Onnela11:Geographic}. The authors observed that communities of 30 or fewer members tended to be geographically tight, with a 100\% increase in the geographic spread of communities occurring as the number of members increased from 30 to 40. This might suggest that communities at the intra-city level, such as those we observe in the place-based social networks under study, are intrinsically constrained in size in order to be sustainable within the city. This idea is related to previous studies that have found an apparent upper limit of around 30 on the maximum number of more intense relationships that a person may be able to maintain, with contacts who are not part of this `inner circle' being communicated with less frequently~\cite{Hill03:Social,Sutcliffe12:Relationships}. In the city-level social networks, one way in which this threshold could manifest is as a limit on the maximum number of people with whom one is able to maintain regular face-to-face contact at a place, reflected in the size of close-knit groups based around places. As we see here, this affects the community structure of the resulting social network.

\section{Conclusions and future work}
In this work, we performed a large-scale investigation into whether places could act as foci for friendships in a social network, in a way consistent with the focused organization theory of social ties. We used a  dataset from the online location-based social network Foursquare to analyze intra-city social networks, and we confirmed that they have the structural properties observed by sociologists and computational social scientists studying real-world social networks. Looking deeper into the spatial characteristics of the networks, we showed that triangles in the social network often have a single common place shared between all three members, and that certain types of places seem more likely to indicate friendship between individuals who meet at those places than others, which fits with the focused organization theory.

We then used our observations to propose a model for social networks in cities based on focused tie organization with places as foci, and showed that the model generates networks with the well-known structural properties of social networks. We demonstrated the importance of the places as foci in our model by analyzing the effect of each individual piece of information on the resulting networks, as well as showing that the model preserves characteristic spatial properties of empirical data.

We intend as future work to investigate how this model of the formation of friendships around certain kinds of places could be exploited in these ways by online location-based social services, for example, in smarter privacy controls based on the semantics of a relationship inferred from types of meeting places, and providing better recommendations for places to visit for groups of friends going out together.

\section{Acknowledgements}
Chlo\"{e} Brown is a recipient of the Google Europe Fellowship in Mobile Computing, and this research is supported in part by this Google Fellowship. We acknowledge the support of EPSRC through grant GALE (EP/K019392).
\bibliography{biblio}
\end{document}